\documentclass{svmult}
\usepackage{epsfig,graphicx} 
\usepackage[bottom]{footmisc}
\usepackage{natbib,url}      
\bibpunct{(}{)}{;}{a}{}{,}   
\def\cite#1{\citealp{#1}}    
\def\authorindex#1{}
\def\figspath{.}  

\begin{document}\newcount\preprintheader\preprintheader=1



\def\thisvolume{these proceedings}

\def\aj{{AJ}}			
\def\araa{{ARA\&A}}		
\def\apj{{ApJ}}			
\def\apjl{{ApJ}}		
\def\apjs{{ApJS}}		
\def\ao{{Appl.\ Optics}} 
\def\apss{{Ap\&SS}}		
\def\aap{{A\&A}}		
\def\aapr{{A\&A~Rev.}}		
\def\aaps{{A\&AS}}		
\def\an{{Astron.\ Nachrichten}}
\def\aspcs{{ASP Conf.\ Ser.}}
\def\assp{{Astrophys.\ \& Space Sci.\ Procs., Springer, Heidelberg}}
\def\azh{{AZh}}			
\def\baas{{BAAS}}		
\def\jrasc{{JRASC}}	
\def\memras{{MmRAS}}		
\def\mnras{{MNRAS}}
\def\nat{{Nat}}		
\def\pra{{Phys.\ Rev.\ A}} 
\def\prb{{Phys.\ Rev.\ B}}		
\def\prc{{Phys.\ Rev.\ C}}		
\def\prd{{Phys.\ Rev.\ D}}		
\def\prl{{Phys.\ Rev.\ Lett.}} 
\def\pasp{{PASP}}
\def\pasj{{PASJ}}		
\def\qjras{{QJRAS}}
\def\science{{Sci}}		
\def\skytel{{S\&T}}		
\def\solphys{{Solar\ Phys.}} 
\def\sovast{{Soviet\ Ast.}}  
\def\ssr{{Space\ Sci.\ Rev.}}
\def\svassp{{Astrophys.\ Space Sci.\ Procs., Springer, Heidelberg}}
\def\zap{{ZAp}}			
\let\astap=\aap
\let\apjlett=\apjl
\let\apjsupp=\apjs
\def\grl{{Geophys.\ Res.\ Lett.}}  
\def\jgr{{J. Geophys.\ Res.}} 

\def\ion#1#2{{\rm #1}\,{\uppercase{#2}}}  
\def\deg{\hbox{$^\circ$}}
\def\sun{\hbox{$\odot$}}
\def\earth{\hbox{$\oplus$}}
\def\la{\mathrel{\hbox{\rlap{\hbox{\lower4pt\hbox{$\sim$}}}\hbox{$<$}}}}
\def\ga{\mathrel{\hbox{\rlap{\hbox{\lower4pt\hbox{$\sim$}}}\hbox{$>$}}}}
\def\sq{\hbox{\rlap{$\sqcap$}$\sqcup$}}
\def\arcmin{\hbox{$^\prime$}}
\def\arcsec{\hbox{$^{\prime\prime}$}}
\def\fd{\hbox{$.\!\!^{\rm d}$}}
\def\fh{\hbox{$.\!\!^{\rm h}$}}
\def\fm{\hbox{$.\!\!^{\rm m}$}}
\def\fs{\hbox{$.\!\!^{\rm s}$}}
\def\fdg{\hbox{$.\!\!^\circ$}}
\def\farcm{\hbox{$.\mkern-4mu^\prime$}}
\def\farcs{\hbox{$.\!\!^{\prime\prime}$}}
\def\fp{\hbox{$.\!\!^{\scriptscriptstyle\rm p}$}}
\def\micron{\hbox{$\mu$m}}
\def\onehalf{\hbox{$\,^1\!/_2$}}	
\def\onethird{\hbox{$\,^1\!/_3$}}
\def\twothirds{\hbox{$\,^2\!/_3$}}
\def\onequarter{\hbox{$\,^1\!/_4$}}
\def\threequarters{\hbox{$\,^3\!/_4$}}
\def\ubv{\hbox{$U\!BV$}}		
\def\ubvr{\hbox{$U\!BV\!R$}}		
\def\ubvri{\hbox{$U\!BV\!RI$}}		
\def\ubvrij{\hbox{$U\!BV\!RI\!J$}}		
\def\ubvrijh{\hbox{$U\!BV\!RI\!J\!H$}}		
\def\ubvrijhk{\hbox{$U\!BV\!RI\!J\!H\!K$}}		
\def\ub{\hbox{$U\!-\!B$}}		
\def\bv{\hbox{$B\!-\!V$}}		
\def\vr{\hbox{$V\!-\!R$}}		
\def\ur{\hbox{$U\!-\!R$}}


\def\labelitemi{{\bf --}}  

\def\rmit#1{{\it #1}}              
\def\rmit#1{{\rm #1}}              
\def\etal{\rmit{et al.}}           
\def\etc{\rmit{etc.}}           
\def\ie{\rmit{i.e.,}}              
\def\eg{\rmit{e.g.,}}              
\def\cf{cf.}                       
\def\viz{\rmit{viz.}}
\def\vs{\rmit{vs.}}

\def\rot{\hbox{\rm rot}}
\def\div{\hbox{\rm div}}
\def\lesssim{\mathrel{\hbox{\rlap{\hbox{\lower4pt\hbox{$\sim$}}}\hbox{$<$}}}}
\def\gtrsim{\mathrel{\hbox{\rlap{\hbox{\lower4pt\hbox{$\sim$}}}\hbox{$>$}}}}
\def\dif{\: {\rm d}}                       
\def\ep{\:{\rm e}^}                        
\def\dash{\hbox{$\,-\,$}}                  
\def\is{\!=\!}                             

\def\starname#1#2{${#1}$\,{\rm {#2}}}  
\def\Teff{\hbox{$T_{\rm eff}$}}   

\def\kms{\hbox{km$\;$s$^{-1}$}}
\def\ms{\hbox{m$\;$s$^{-1}$}}
\def\Mxcm{\hbox{Mx\,cm$^{-2}$}}    

\def\Bapp{\hbox{$B_{\rm app}$}}    

\def\komega{($k, \omega$)}                 
\def\kf{($k_h,f$)}                         
\def\VminI{\hbox{$V\!\!-\!\!I$}}           
\def\IminI{\hbox{$I\!\!-\!\!I$}}           
\def\VminV{\hbox{$V\!\!-\!\!V$}}           
\def\Xt{\hbox{$X\!\!-\!t$}}                

\def\level #1 #2#3#4{$#1 \: ^{#2} \mbox{#3} ^{#4}$}   

\def\specchar#1{\uppercase{#1}}    
\def\AlI{\mbox{Al\,\specchar{i}}}  
\def\BI{\mbox{B\,\specchar{i}}} 
\def\BII{\mbox{B\,\specchar{ii}}}  
\def\BaI{\mbox{Ba\,\specchar{i}}}  
\def\BaII{\mbox{Ba\,\specchar{ii}}} 
\def\CI{\mbox{C\,\specchar{i}}} 
\def\CII{\mbox{C\,\specchar{ii}}} 
\def\CIII{\mbox{C\,\specchar{iii}}} 
\def\CIV{\mbox{C\,\specchar{iv}}} 
\def\CaI{\mbox{Ca\,\specchar{i}}} 
\def\CaII{\mbox{Ca\,\specchar{ii}}} 
\def\CaIII{\mbox{Ca\,\specchar{iii}}} 
\def\CoI{\mbox{Co\,\specchar{i}}} 
\def\CrI{\mbox{Cr\,\specchar{i}}} 
\def\CriI{\mbox{Cr\,\specchar{ii}}} 
\def\CsI{\mbox{Cs\,\specchar{i}}} 
\def\CsII{\mbox{Cs\,\specchar{ii}}} 
\def\CuI{\mbox{Cu\,\specchar{i}}} 
\def\FeI{\mbox{Fe\,\specchar{i}}} 
\def\FeII{\mbox{Fe\,\specchar{ii}}} 
\def\FeIX{\mbox{Fe\,\specchar{ix}}}
\def\FeX{\mbox{Fe\,\specchar{x}}}
\def\FeXVI{\mbox{Fe\,\specchar{xvi}}}
\def\FrI{\mbox{Fr\,\specchar{i}}}
\def\HI{\mbox{H\,\specchar{i}}} 
\def\HII{\mbox{H\,\specchar{ii}}} 
\def\Hmin{\hbox{\rmH$^{^{_{\scriptstyle -}}}$}}      
\def\Hemin{\hbox{{\rm He}$^{^{_{\scriptstyle -}}}$}} 
\def\HeI{\mbox{He\,\specchar{i}}} 
\def\HeII{\mbox{He\,\specchar{ii}}} 
\def\HeIII{\mbox{He\,\specchar{iii}}} 
\def\KI{\mbox{K\,\specchar{i}}} 
\def\KII{\mbox{K\,\specchar{ii}}} 
\def\KIII{\mbox{K\,\specchar{iii}}} 
\def\LiI{\mbox{Li\,\specchar{i}}} 
\def\LiII{\mbox{Li\,\specchar{ii}}} 
\def\LiIII{\mbox{Li\,\specchar{iii}}} 
\def\MgI{\mbox{Mg\,\specchar{i}}} 
\def\MgII{\mbox{Mg\,\specchar{ii}}} 
\def\MgIII{\mbox{Mg\,\specchar{iii}}} 
\def\MnI{\mbox{Mn\,\specchar{i}}} 
\def\NI{\mbox{N\,\specchar{i}}}
\def\NIV{\mbox{N\,\specchar{iv}}}
\def\NaI{\mbox{Na\,\specchar{i}}}
\def\NaII{\mbox{Na\,\specchar{ii}}}
\def\NaIII{\mbox{Na\,\specchar{iii}}}
\def\NeVIII{\mbox{Ne\,\specchar{viii}}} 
\def\NiI{\mbox{Ni\,\specchar{i}}} 
\def\NiII{\mbox{Ni\,\specchar{ii}}}
\def\NiIII{\mbox{Ni\,\specchar{iii}}} 
\def\OI{\mbox{O\,\specchar{i}}} 
\def\OVI{\mbox{O\,\specchar{vi}}}
\def\RbI{\mbox{Rb\,\specchar{i}}} 
\def\SII{\mbox{S\,\specchar{ii}}} 
\def\SiI{\mbox{Si\,\specchar{i}}} 
\def\SiII{\mbox{Si\,\specchar{ii}}} 
\def\SrI{\mbox{Sr\,\specchar{i}}}
\def\SrII{\mbox{Sr\,\specchar{ii}}}
\def\TiI{\mbox{Ti\,\specchar{i}}} 
\def\TiII{\mbox{Ti\,\specchar{ii}}} 
\def\TiIII{\mbox{Ti\,\specchar{iii}}} 
\def\TiIV{\mbox{Ti\,\specchar{iv}}} 
\def\VI{\mbox{V\,\specchar{i}}} 
\def\HtwoO{\mbox{H$_2$O}}        
\def\Otwo{\mbox{O$_2$}}          

\def\Halpha{\mbox{H\hspace{0.1ex}$\alpha$}} 
\def\Ha{\mbox{H\hspace{0.2ex}$\alpha$}}
\def\Hbeta{\mbox{H\hspace{0.2ex}$\beta$}}
\def\Hgamma{\mbox{H\hspace{0.2ex}$\gamma$}}
\def\Hdelta{\mbox{H\hspace{0.2ex}$\delta$}}
\def\Hepsilon{\mbox{H\hspace{0.2ex}$\epsilon$}}
\def\Hzeta{\mbox{H\hspace{0.2ex}$\zeta$}}
\def\Lyalpha{\mbox{Ly$\hspace{0.2ex}\alpha$}}
\def\Lybeta{\mbox{Ly$\hspace{0.2ex}\beta$}}
\def\Lygamma{\mbox{Ly$\hspace{0.2ex}\gamma$}}
\def\Lycont{\mbox{Ly\hspace{0.2ex}{\small cont}}}
\def\Baalpha{\mbox{Ba$\hspace{0.2ex}\alpha$}}
\def\Babeta{\mbox{Ba$\hspace{0.2ex}\beta$}}
\def\Bacont{\mbox{Ba\hspace{0.2ex}{\small cont}}}
\def\Paalpha{\mbox{Pa$\hspace{0.2ex}\alpha$}}
\def\Bralpha{\mbox{Br$\hspace{0.2ex}\alpha$}}

\def\NaD{\mbox{Na\,\specchar{i}\,D}}    
\def\NaDone{\mbox{Na\,\specchar{i}\,\,D$_1$}}
\def\NaDtwo{\mbox{Na\,\specchar{i}\,\,D$_2$}}
\def\NaID{\mbox{Na\,\specchar{i}\,\,D}}
\def\NaIDone{\mbox{Na\,\specchar{i}\,\,D$_1$}}
\def\NaIDtwo{\mbox{Na\,\specchar{i}\,\,D$_2$}}
\def\Done{\mbox{D$_1$}}
\def\Dtwo{\mbox{D$_2$}}

\def\Mgbone{\mbox{Mg\,\specchar{i}\,b$_1$}}
\def\Mgbtwo{\mbox{Mg\,\specchar{i}\,b$_2$}}
\def\Mgbthree{\mbox{Mg\,\specchar{i}\,b$_3$}}
\def\MgIb{\mbox{Mg\,\specchar{i}\,b}}
\def\MgIbone{\mbox{Mg\,\specchar{i}\,b$_1$}}
\def\MgIbtwo{\mbox{Mg\,\specchar{i}\,b$_2$}}
\def\MgIbthree{\mbox{Mg\,\specchar{i}\,b$_3$}}

\def\CaIIK{\mbox{Ca\,\specchar{ii}\,K}}       
\def\CaIIH{\mbox{Ca\,\specchar{ii}\,H}}
\def\CaIIHK{\mbox{Ca\,\specchar{ii}\,H\,\&\,K}}
\def\HK{\mbox{H\,\&\,K}}
\def\Kthree{\mbox{K$_3$}}      
\def\Hthree{\mbox{H$_3$}}
\def\Ktwo{\mbox{K$_2$}}
\def\Htwo{\mbox{H$_2$}}
\def\Kone{\mbox{K$_1$}}     
\def\Hone{\mbox{H$_1$}}     
\def\KtwoV{\mbox{K$_{2V}$}}
\def\KtwoR{\mbox{K$_{2R}$}}
\def\KoneV{\mbox{K$_{1V}$}}
\def\KoneR{\mbox{K$_{1R}$}}
\def\HtwoV{\mbox{H$_{2V}$}}
\def\HtwoR{\mbox{H$_{2R}$}}
\def\HoneV{\mbox{H$_{1V}$}}
\def\HoneR{\mbox{H$_{1R}$}}

\def\hk{\mbox{h\,\&\,k}}
\def\kthree{\mbox{k$_3$}}    
\def\hthree{\mbox{h$_3$}}
\def\ktwo{\mbox{k$_2$}}
\def\htwo{\mbox{h$_2$}}
\def\kone{\mbox{k$_1$}}     
\def\hone{\mbox{h$_1$}}     
\def\ktwoV{\mbox{k$_{2V}$}}
\def\ktwoR{\mbox{k$_{2R}$}}
\def\koneV{\mbox{k$_{1V}$}}
\def\koneR{\mbox{k$_{1R}$}}
\def\htwoV{\mbox{h$_{2V}$}}
\def\htwoR{\mbox{h$_{2R}$}}
\def\honeV{\mbox{h$_{1V}$}}
\def\honeR{\mbox{h$_{1R}$}}

\ifnum\preprintheader=1     
\makeatletter  
\def\@maketitle{\newpage
\markboth{}{}%
  {\mbox{} \vspace*{-8ex} \par 
   \em \footnotesize To appear in ``Magnetic Coupling between the Interior 
       and the Atmosphere of the Sun'', eds. S.~S.~Hasan and R.~J.~Rutten, 
       Astrophysics and Space Science Proceedings, Springer-Verlag, 
       Heidelberg, Berlin, 2009.} \vspace*{-5ex} \par
 \def\lastand{\ifnum\value{@inst}=2\relax
                 \unskip{} \andname\
              \else
                 \unskip \lastandname\
              \fi}%
 \def\and{\stepcounter{@auth}\relax
          \ifnum\value{@auth}=\value{@inst}%
             \lastand
          \else
             \unskip,
          \fi}%
  \raggedright
 {\Large \bfseries\boldmath
  \pretolerance=10000
  \let\\=\newline
  \raggedright
  \hyphenpenalty \@M
  \interlinepenalty \@M
  \if@numart
     \chap@hangfrom{}
  \else
     \chap@hangfrom{\thechapter\thechapterend\hskip\betweenumberspace}
  \fi
  \ignorespaces
  \@title \par}\vskip .8cm
\if!\@subtitle!\else {\large \bfseries\boldmath
  \vskip -.65cm
  \pretolerance=10000
  \@subtitle \par}\vskip .8cm\fi
 \setbox0=\vbox{\setcounter{@auth}{1}\def\and{\stepcounter{@auth}}%
 \def\thanks##1{}\@author}%
 \global\value{@inst}=\value{@auth}%
 \global\value{auco}=\value{@auth}%
 \setcounter{@auth}{1}%
{\lineskip .5em
\noindent\ignorespaces
\@author\vskip.35cm}
 {\small\institutename\par}
 \ifdim\pagetotal>157\p@
     \vskip 11\p@
 \else
     \@tempdima=168\p@\advance\@tempdima by-\pagetotal
     \vskip\@tempdima
 \fi
}
\makeatother     
\fi


\title*{Measuring the Hidden Aspects of Solar Magnetism}

\author{J. O. Stenflo}
\authorindex{Stenflo, J. O.}
\institute{Institute of Astronomy, ETH Zurich}
\maketitle

\setcounter{footnote}{0}  

\begin{abstract}
2008 marks the 100th anniversary of the discovery of astrophysical magnetic fields, when George Ellery Hale recorded the Zeeman splitting of spectral lines in sunspots. With the introduction of Babcock's photoelectric magnetograph it soon became clear that the Sun's magnetic field outside sunspots is extremely structured. The field strengths that were measured were found to get larger when the spatial resolution was improved. It was therefore necessary to come up with methods to go beyond the spatial resolution limit and diagnose the intrinsic magnetic-field properties without dependence on the quality of the telescope used. The line-ratio technique that was developed in the early 1970s revealed a picture where most flux that we see in magnetograms originates in highly bundled, kG fields with a tiny volume filling factor. This led to interpretations in terms of discrete, strong-field magnetic flux tubes embedded in a rather field-free medium, and a whole industry of flux tube models at increasing levels of sophistication. This magnetic-field paradigm has now been shattered with the advent of high-precision imaging polarimeters that allow us to apply the so-called ``Second Solar Spectrum'' to diagnose aspects of solar magnetism that have been hidden to Zeeman diagnostics. It is found that the bulk of the photospheric volume is seething with intermediately strong, tangled fields. In the new paradigm the field behaves like a fractal with a high degree of self-similarity, spanning about 8 orders of magnitude in scale size, down to scales of order 10\,m. 
\end{abstract}


\section{The Zeeman effect as a window to cosmic magnetism}\label{stenflo-sec:zeeman}

\begin{figure}
 \includegraphics[angle=-90.,width=1.\textwidth]{\figspath/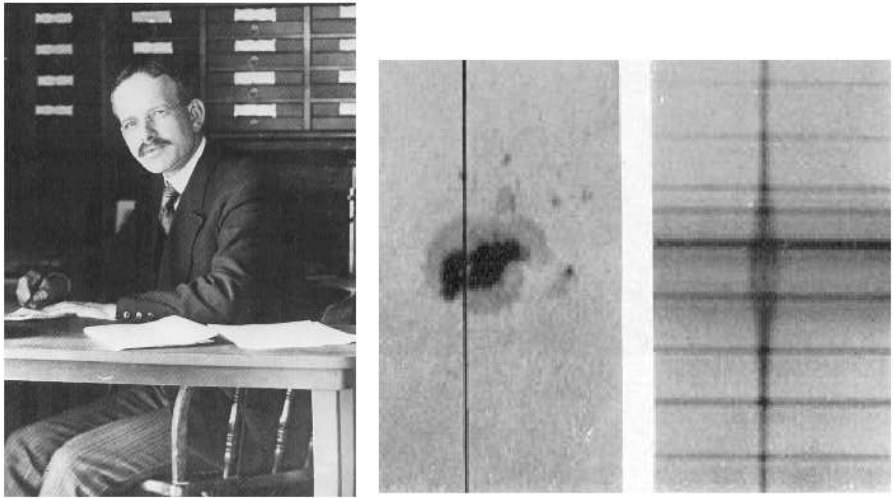}
 \caption[]{\label{stenflo-fig:hale}George Ellery Hale in 1905 in his
 office at the Mount Wilson Observatory (left) and his discovery of
 the Zeeman splitting in sunspots (right).}
\end{figure}

2008 marks the 100th anniversary of the discovery of magnetic fields outside the Earth (cf.~Fig.~\ref{stenflo-fig:hale}). George Ellery Hale had suspected that the Sun might be a magnetized sphere from the appearance of the solar corona seen at total solar eclipses, and from the structure of H$\alpha$ fibrils around sunspots, which was reminiscent of iron files in a magnetic field. The proof came when Hale placed the spectrograph slit in the solar tower of his newly constructed Mt. Wilson Observatory across a sunspot and the splitting of the spectral lines was revealed (Hale 1908). The Dutch physicist Pieter Zeeman had discovered such splitting in the laboratory the decade before, in 1896, when an external magnetic field was applied to the gas in which the spectral lines were formed. The frequency of Larmor precession of the atomic dipoles around the field mixes with the atomic resonance frequencies, resulting in the splitting of these frequencies. Since the magnetic field breaks the spatial symmetry the split line components get polarized in ways that depend on the orientation of the magnetic field vector relative to the line of sight. 

Often the splitting is too small and subtle to measure by itself except in sunspots. Instead it is the polarization effects that are the telltale signature of the Zeeman effect and the magnetic field. While the line-of-sight component of the magnetic field gives rise to circular polarization (longitudinal Zeeman effect), the perpendicular component causes linear polarization (transverse Zeeman effect). Thus one can in principle measure the full magnetic field vector (strength and orientation) by recording the full state of polarization in a spectral line, the full Stokes vector, with the four Stokes parameters $I$, $Q$, $U$, $V$, representing intensity, two states of linear polarization (differing in orientation by $45^\circ$), and circular polarization, respectively. Detailed treatments can be found in several monographs (\cite{stenflo-book94}; \cite{stenflo-dti03}; \cite{stenflo-lanlan04}). 

The Zeeman effect, which outside terrestrial laboratories was first seen in sunspots, opens a window to the exploration of cosmic magnetism. As usual the Sun provides us with a laboratory that serves as a test bed for the exploration of various new physical processes before they can be applied to the investigation of stars and galaxies elsewhere in the universe. Our increasing empirical knowledge about the Sun's magnetism has helped guide the development and understanding of various theoretical tools, like plasma physics and magnetohydrodynamics. The experimental tool is spectro-polarimetry, which needs the Zeeman effect (and more recently also the Hanle effect, see below) as an interpretational tool to connect theory and observation. 

Outside sunspots the polarization signals of the transverse Zeeman effect are much smaller than those of the longitudinal Zeeman effect. For weak fields the linear polarization from the transverse Zeeman effect is approximately proportional to the square of the transverse field strength rather than in linear proportion, and it is limited by a $180^\circ$ ambiguity. In contrast, the circular polarization is easy to measure, and to first order it is proportional to the line-of-sight component of the field, with sign. Therefore magnetic-field measurements have been dominated by recordings of the circular polarization due to the longitudinal Zeeman effect. The breakthrough in these measurements came with the introduction by Babcock of the photoelectric magnetograph (\cite{stenflo-babcock53}). Soon afterwards, full-disk magnetograms (maps of the circular polarization) were being produced on a regular basis, forming a unique data base for the understanding of stellar magnetism and dynamos.


\section{Emergence of the flux tube paradigm}\label{stenflo-sec:fluxtube}
When directly resolved magnetic-field observations are not available, like for magnetic Ap-type stars, one usually makes models assuming that the star has a dipole or low-degree multipolar field. The solar magnetograms however showed the Sun's field to be highly structured. It was found that the measured field strength increases with the angular resolution of the instrument used (\cite{stenflo-s66}). As the measured field strength also depended on the spectral line used, many believed that this was a calibration problem that could be solved by a coordinated campaign, organized by an IAU committee, to record the same regions on the Sun with different instruments. 

It was only with the introduction of the line-ratio technique (Stenflo 1973) that the cause for this apparent ``calibration problem'' could be found. The magnetic flux is highly intermittent, with most of the flux concentrated in elements that were far smaller than the available spatial resolution. The magnetograph calibration (conversion of measured polarization to field strength) was based on the shape of the spatially averaged line profile and the assumption of weak fields (linear relation between polarization and field strength). The average line profile is however not representative of the line formation conditions within the flux concentrations, and also the weak-field approximation is not valid there (we have ``Zeeman saturation''), since the concentrated fields are intrinsically strong. Inside the strong-field regions the thermodynamic conditions are very different from the rest of the atmosphere, which leads to temperature-induced line weakenings. 

\begin{figure}
 \includegraphics[angle=-90.,width=1.\textwidth]{\figspath/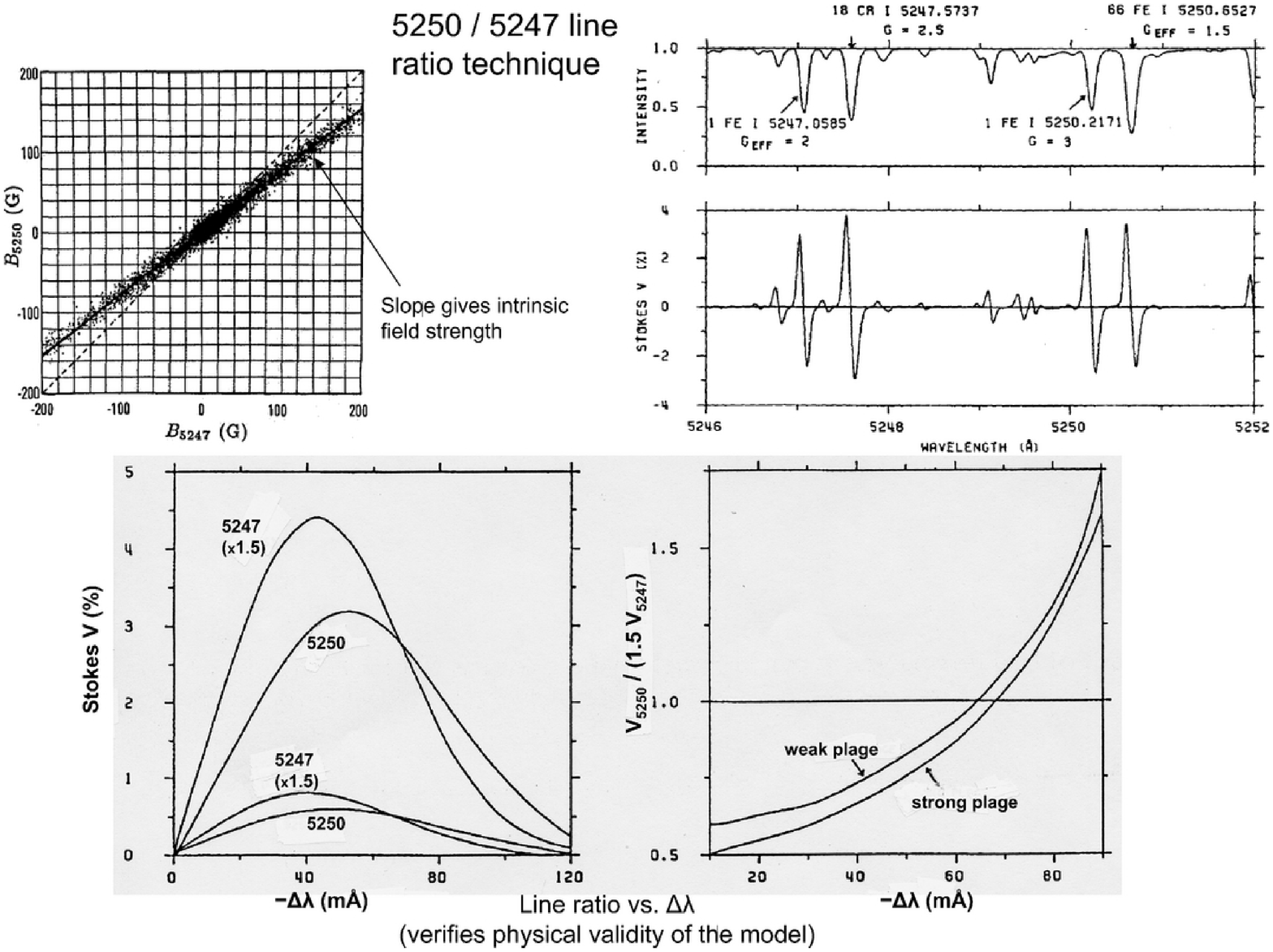}
  \caption[]{\label{stenflo-fig:lineratio}Illustration of the various aspects of the 5250/5247 line ratio technique (\cite{stenflo-s73}). The linear slope in the diagram to upper left (from \cite{stenflo-fs78}) determines the differential Zeeman saturation, from which the intrinsic field strength can be found. The portion of the FTS Stokes $V$ spectrum to upper right, from \citet{stenflo-setal84},  shows that the amplitudes of the 5250 and 5247 iron lines are not in proportion to their Land\'e\ factors, but are closer to 1:1. In the bottom diagram, from \citet{stenflo-sh85}, the Stokes $V$ profiles and line ratios are plotted as functions of wavelength distance from line center. This profile behavior verifies that the line difference is really due to differential Zeeman saturation.}
\end{figure}

The magnitude of the line-weakening and Zeeman saturation effects vary from line to line, which leads to the noticed dependence of the field-strength values on the spectral line used. This effect cannot be calibrated away, since the line-formation properties in the flux concentrations are not accessible to direct observations when they are not resolved. A further effect is that different lines are formed at different atmospheric heights, and the field expands and weakens with height. All these effects contribute jointly in an entangled way to the ``calibration error''. The line-ratio technique was introduced to untangle them. It is described in Fig.~\ref{stenflo-fig:lineratio}. 

The trick is to use a combination of lines, for which all the various entangled factors are identical, except one. Thus it was possible to isolate the Zeeman saturation (non-linearity) effect from all the thermodynamic and line formation effects by choosing the line pair Fe\,{\sc i} 5250.22 and 5247.06\,\AA. Both these lines belong to multiplet no.~1 of iron, have the same line strength and excitation potential, and therefore have identical thermodynamic response and line-formation properties. The only significant difference between them is their Land\'e\ factors, which are 3.0 and 2.0, respectively. No other line combination has since been found, which can so cleanly isolate the Zeeman saturation effect from the other effects. 

If we were in the linear, weak-field regime, the circular polarization measured in the two lines should scale in proportion to their Land\'e\ factors, but as the field strength increases, the deviation from this ratio increases ({\it differential Zeeman saturation}). Thus the circular-polarization line ratio is a direct measure of the intrinsic field strength. The observed ratio showed that the intrinsic field strength was 1-2\,kG at the quiet-sun disk center, although the apparent magnetograph field strengths there were only a few G, a discrepancy of 2-3 orders of magnitude (\cite{stenflo-s73})\,! 

A further surprising result was that there seemed to be no dependence of the intrinsic field strength on the apparent field strength (which in a first approximation represents magnetic flux divided by the spatial resolution element). This property is seen in the scatter-plot diagram to the upper left in Fig.~\ref{stenflo-fig:lineratio} (from \cite{stenflo-fs78}). The line ratio or differential Zeeman saturation is represented by the {\it slope} in the diagram (in comparison with the $45^\circ$ slope that represents the case without Zeeman saturation). There is no indication that the slope changes as we go from smaller to larger apparent field strengths. A statistical analysis led to the conclusion that more than 90\,\%\ of the photospheric flux (that is ``seen'' by the magnetographs with the resolution of a few arcsec that was used then) is in strong-field form (\cite{stenflo-hs72}; \cite{stenflo-fs72}; \cite{stenflo-book94}), and that strong-field flux elements have ``unique'' internal properties, meaning that the statistical spread in their field strengths and thermodynamic properties was small and not dependent on the amount of flux in the region. Thus active-region plages and the quiet-sun network gave very similar intrinsic field strengths. 

These findings lay the foundation for the validity of the 2-component model that was used as the interpretational tool: One ``magnetic'' component with a certain filling factor (fractional area of the resolution element covered), which was the source of all the circular-polarization signals seen in magnetograms, and another component, which was called ``non-magnetic'', since it did not contribute anything to the magnetograms. The line-ratio method showed that the field strength of the magnetic component was nearly independent of the magnetic filling factor, which could vary by orders of magnitude (but had typical values of order 1\,\%\ on the quiet Sun). 

The empirical foundation for the 2-component model was further strengthened by the powerful Stokes $V$ multi-line profile constraints provided by FTS (Fourier transform spectrometer) polarimetry (\cite{stenflo-setal84}), and by the use of the larger Zeeman splitting in the near infrared (cf.~\cite{stenflo-retal92}). 

This empirical scenario found its theoretical counterpart in the concept of strong-field magnetic flux tubes embedded in field-free surroundings (\cite{stenflo-spruit76}). Semi-empirical flux tube models of increasing sophistication could be built, in particular thanks to the powerful observational constraints provided by the FTS Stokes $V$ spectra (cf.~\cite{stenflo-sol93}). In these models the observational constraints were combined with the MHD constraints that included the self-consistent expansion of the flux tubes with height in a numerically specified atmosphere with pressure balance. 

With these successes the unphysical nature of the 2-component model tended to be forgotten, according to which something like 99\,\%\ of the photosphere was ``non-magnetic''. In the electrically highly conducting solar plasma the concept of such a field-free volume is non-sensical. When the 2-component model was introduced nearly four decades ago the introduction of a ``non-magnetic'' component was done for the sake of mathematical simplicity, with the purpose of isolating the properties of the magnetic component, but not with the intention of making a statement about the intrinsic nature of the ``non-magnetic'' component. Since the longitudinal Zeeman effect was ``blind'' to this component (as it did not contribute to anything in the magnetograms), the quest began to find another diagnostic tool to access its hidden magnetic properties, to find a diagnostic window to the aspect of solar magnetism that represents 99\,\%\ of the photosphere. This window was found through the Hanle effect.


\section{The Hanle effect as a window to the hidden fields}\label{stenflo-sec:hanle}
The circular polarization from the longitudinal Zeeman effect is to first order proportional to the net magnetic flux through the angular resolution element. If the magnetic field has mixed-polarity fields inside the resolution element with equal total amounts of positive and negative polarity flux, the net flux and therefore also the net circular polarization is zero. Although the strength and magnetic energy density of such a tangled field can be arbitrarily high, it is invisible to the longitudinal Zeeman effect as long as the individual flux elements are not resolved. 

If this were merely a matter of insufficient angular resolution, one might hope that this tangled field could be mapped by magnetograms in some future. However, even if we would have infinite angular resolution, the cancellation problem of the opposite polarities would not go away, since the spatial resolution {\it along the line of sight} is ultimately limited by the thickness of the line-forming layer, which is of order 100\,km in the photosphere (the photon mean free path). For optically thin magnetic elements with opposite polarities along the line of sight the cancellation effect remains, regardless of the angular resolution. 

The task therefore becomes to find a physical mechanism that is not subject to these cancellation effects. Magnetic line broadening from the Zeeman effect is one such mechanism, since it scales with the square of the field strength, the magnetic energy, and therefore is of one ``sign'', in contrast to the circular polarization. Since however these effects are tiny, and many other factors affect the width of spectral lines, only a 1-$\sigma$ upper limit of about 100\,G could be set for the tangled field from a statistical study of 400 unblended Fe\,{\sc i} lines (\cite{stenflo-sl77}). In contrast, the Hanle effect is sensitive to much weaker tangled fields.  

\begin{figure}
\centering
 \includegraphics[width=0.7\textwidth]{\figspath/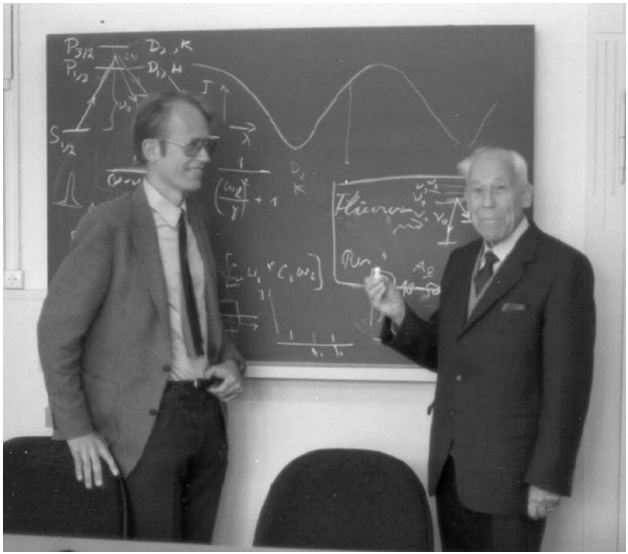}
  \caption[]{\label{stenflo-fig:hanle}Wilhelm Hanle (right) visits ETH Zurich in 1983 on the occasion of the 60th anniversary of his effect.}
\end{figure}

In contrast to the Zeeman effect, the Hanle effect is a coherence phenomenon that only occurs when coherent scattering contributes to the formation of the spectral line. It was discovered in G\"ottingen in 1923 by Wilhelm Hanle and played a significant role in the conceptual development of quantum mechanics, since it demonstrated explicitly the fundamental concept of the coherent superposition of quantum states (later sometimes called ``Schr\"odinger cats''). 

Coherent scattering polarizes the light. The term Hanle effect covers all the magnetic-field modifications of this scattering polarization. In the absence of magnetic fields the magnetic $m$ substates are degenerate (coherently superposed). A magnetic field breaks the spatial symmetry and lifts the degeneracy, thereby causing partial decoherence. One can also speak of quantum interferences between the $m$ states. For details, see \citet{stenflo-mor91}; \citet{stenflo-book94}; \citet{stenflo-lanlan04}. 

A good intuitive understanding of the Hanle effect can be obtained with the help of the classical oscillator model. The incident radiation induces dipole oscillations in the transverse plane (perpendicular to the incident beam). For a $90^\circ$ scattering angle the plane in which the oscillations take place is viewed from the side and due to this projection appear as 1-D oscillations. The scattered radiation therefore gets 100\,\%\ linearly polarized perpendicular to the scattering plane. 

For scattering polarization to occur one needs anisotropic radiative excitation. For a spherically symmetric Sun (when we neglect local inhomogeneities), the anisotropy is a consequence of the limb darkening, which implies that the illumination of a scattering particle inside the atmosphere occurs more in the vertical direction from below than from the sides. In the hypothetical case of extreme limb darkening, when all illumination is in the vertical direction, we would have $90^\circ$ scattering at the extreme limb. The scattering angle decreases towards zero when we move towards disk center, where for symmetry reasons the scattering polarization (in the non-magnetic case) is zero. Since the scattering polarization gets larger as we approach the limb, most scattering and Hanle-effect observations are performed on the disk relatively close to the limb, with the spectrograph slit parallel to the nearest limb. The non-magnetic scattering polarization is then expected to be oriented along the slit direction, which we in our Stokes vector representations define as the positive Stokes $Q$ direction. Stokes $U$ then represents polarization oriented at $45^\circ$ to the slit. 

\begin{figure}
 \includegraphics[angle=-90.,width=1.\textwidth]{\figspath/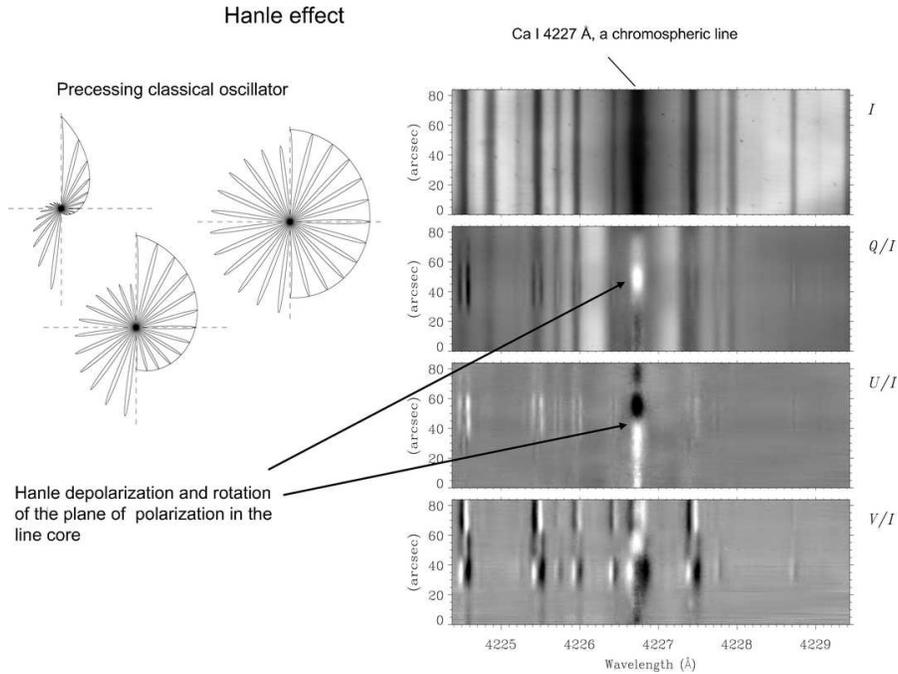}
  \caption[]{\label{stenflo-fig:hanle2}Left diagrams: Rosette patterns of a classical oscillator in a magnetic field oriented along the line of sight, illustrating the Hanle depolarization and rotation effects. Right diagram: Spectral image of the Stokes vector (the four Stokes parameters in terms of intensity $I$ and the fractional polarizations $Q/I$, $U/I$, and $V/I$) recorded with the spectrograph slit across a moderately magnetic region 5\,arcsec inside and parallel to the solar limb. The Hanle signatures appear in Stokes $Q$ and $U$ in the core of the Ca\,{\sc i} 4227\,\AA\ line, while the surrounding lines exhibit the characteristic signatures of the transverse Zeeman effect. In Stokes $V$ all the lines show the anti-symmetric signatures of the longitundinal Zeeman effect.}
\end{figure}

Let us now introduce a magnetic field along the scattering direction. The damped oscillator is then subject to Larmor precession around the magnetic field vector, which results in the Rosette patterns illustrated in Fig.~\ref{stenflo-fig:hanle2}. The pattern gets tilted and more randomized as the field strength increases (from the left to the right Rosette diagram in the figure). The line profile and polarization properties are obtained from Fourier transformations of the Rosette patterns. 

The magnetic field has two main effects on the polarization of the scattered radiation: (1) Depolarization, since the precession randomizes the orientations of the oscillating dipoles. In terms of the Stokes parameters, this corresponds to a reduction of the $Q/I$ amplitudes. (2) Rotation of the plane of linear polarization, since the net effect of the precession is a skewed or tilted oscillation pattern. This corresponds to the creation of signatures in Stokes $U/I$, which can be of either sign, depending on the sense of rotation (orientation of the field vector). The magnitudes of these two effects depend on the competition between the Larmor precession rate and the damping rate, or, equivalently, the ratio between the Zeeman splitting and the damping width of the line. In contrast, the polarization caused by the ordinary Zeeman effect depends on the ratio between the Zeeman splitting and the Doppler width of the line. Since the damping width is smaller by typically a factor of 30 than the Doppler width, the Hanle effect is sensitive to much weaker fields than the Zeeman effect. Equally important, the two effects have different symmetry properties and therefore respond to magnetic fields in highly complementary ways. 

Assume for instance that we are observing a magnetic field that is tangled on subresolution scales, such that there is no net magnetic flux when one averages over the spatial resolution element due to cancellation of the contributions of opposite signs. Such a magnetic field gives no observable signatures in the circular polarization (longitudinal Zeeman effect, on which solar magnetograms are based) or in the Hanle rotation (Stokes $U/I$) due to cancellations of the opposite signs. In contrast, the Hanle depolarization is not subject to such cancellations, since it has only one ``sign'' (depolarization), regardless of the field direction. The Hanle depolarization therefore opens a diagnostic window to such a subresolution, tangled field (\cite{stenflo-s82}).


\section{The ``standard model'' and its shortcomings}\label{sec:standard}

\begin{figure}
 \includegraphics[angle=-90.,width=1.\textwidth]{\figspath/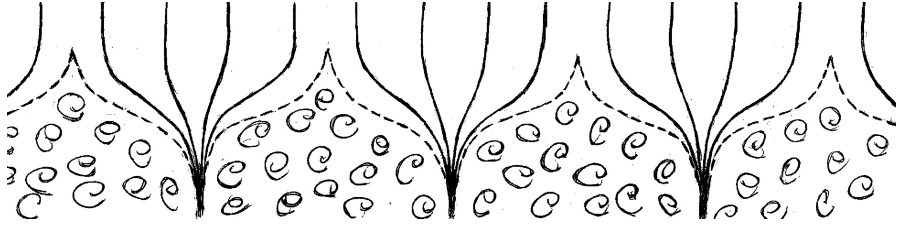}
  \caption[]{\label{stenflo-fig:fluxtube}Standard model of quiet-sun solar magnetism (here illustrated for a region where the different flux tubes have the same polarity). The atmosphere is described in terms of two components, one representing the flux tubes, which contribute to the Zeeman effect, the other component representing the tangled field in between, which contributes to the Hanle effect.}
\end{figure}

The ``standard model'' that has emerged from Zeeman and Hanle observations of the quiet Sun, and which is illustrated in Fig.~\ref{stenflo-fig:fluxtube}, refers to the magnetic-field structuring in the spatially unresolved domain. Only recently, with advances in angular resolution, are we beginning to resolve individual flux tubes, but in general their existence and properties have only been infered from indirect techniques (line-ratio method, FTS Stokes $V$ spectra, Stokes $V$ line profiles in the near infrared). Since the fields are not resolved, all such indirect techniques must be based on interpretative models. 

The dominating interpretative model in the past has been a 2-component model, consisting of (1) the flux tube component, which is responsible for practically all the magnetic flux that is seen in solar magnetograms, and (2) the ``turbulent'' component in between the flux tubes, with tangled fields of mixed polarities on subresolution scales, which are invisible to the Zeeman effect. The filling factor of the flux tube component is of order 1\,\%\ in the quiet solar photosphere, which implies that the turbulent component represents 99\,\%\ of the photospheric volume. Due to the exponential pressure drop with height the flux tubes expand to reach a filling factor of 100\,\%\ in the corona. 

The question about the strength of the volume-filling ``turbulent'' field representing 99\,\%\ of the photosphere could be given an answer from observations of the Hanle depolarization of the scattering polarization, in particular with the Sr\,{\sc i} 4607\,\AA\ line. Since with one such line we only have one observable (the amount of Hanle depolarization), the interpretative model could not have more than one free parameter. The natural choice of 1-parameter model that was adopted in the initial interpretations of the Hanle data was in terms of a tangled field consisting of optically thin elements with a random, isotropic distribution of the magnetic field vectors and a single-valued field strength (\cite{stenflo-s82}). Detailed radiative-transfer modelling of the Sr\,{\sc i} 4607\,\AA\ observations (\cite{stenflo-faurob93}; \cite{stenflo-faurobetal95}) gave values of typically 30\,G, but more recent applications of 3-D polarized radiative transfer for much more realistic model atmospheres generated by hydrodynamic simulations of granular convection give field strengths of about 60\,G, twice as large (\cite{stenflo-trujetal04}). 

The dualistic nature of the world that is represented by this ``standard model'' is however much an artefact of having two mutually almost exclusive diagnostic tools at our disposal. The Zeeman effect is blind to the turbulent fields due to flux cancellation. The Hanle effect is blind to the flux tube fields for several reasons: ({\it i}) With filling factors of order 1\,\%\ only, the flux tube contribution to the Hanle depolarization is insignificant. ({\it ii}) The Hanle effect is insensitive to vertical fields (for symmetry reasons, when the illumination is axially symmetric around the field vector), and the flux tubes tend to be vertical because of buoyancy. ({\it iii}) The Hanle effect saturates for the strong fields in the flux tubes. 

We always see a filtered version of the real world, filtered by our diagnostic tools in combination with the interpretational models (analytical tools) used. Thus, when we put on our ``Zeeman goggles'' we see a magnetic world governed by flux tubes, while when we put on our ``Hanle goggles'', we see a world of tangled or turbulent fields. We should however not forget that these are merely idealized aspects of the real world, shaped by our models. Instead of having the dichotomy of two discrete components, the real world should rather be described in terms of continuous probability density functions (PDFs), as indicated by the theory of magnetoconvection and by numerical simulations (\cite{stenflo-cattaneo99}; \cite{stenflo-nordlund90}). Moreover, exploration of the magnetic pattern on the spatially resolved scales indicates a high degree of self-similarity that is characteristic of a fractal (\cite{stenflo-sh02}; \cite{stenflo-jetal03}). 

When \citet{stenflo-trujetal04} used an interpretational model based on a realistic PDF rather than a single-valued field strength, their 3-D modelling of the Sr\,{\sc i} 4607\,\AA\ observations gave substantially higher average field strengths (in excess of 100\,G) as compared with the single-valued model. This suggests that the hidden, turbulent field contains a magnetic energy density that may be of significance for the overall energy balance of the solar atmosphere. The question whether or not the magnetic energy dominates the energy balance remains unanswered due to the current model dependence of these interpretations.


\section{The Second Solar Spectrum and solar magnetism}\label{sec:sss}
The term Hanle effect stands for the magnetic-field modifications of the scattering polarization. The Sun's spectrum is linearly polarized since coherent scattering contributes to the formation of the spectrum (like the polarization of the blue sky by Rayleigh scattering at terrestrial molecules). Due to the small anisotropy of the radiation field in the solar atmosphere and the competing non-polarizing opacity sources, the amplitudes of the scattering polarization signals are small, of order 0.01 -- 1\,\%\ near the limb, varying from line to line. Although a number of the polarized line profiles could be revealed in early surveys of the linear polarization (\cite{stenflo-setal83a,stenflo-setal83b}), it was only with the advent of highly sensitive imaging polarimeters that the rich spectral world of scattering polarization became fully accessible to observation. The breakthrough came with the implementation in 1994 of the ZIMPOL (Zurich Imaging Polarimeter) technology, which allowed imaging spectro-polarimetry with a precision of $10^{-5}$ in the degree of polarization (\cite{stenflo-povel95,stenflo-povel01}; \cite{stenflo-gandetal04}). At this level of sensitivity everything is polarized, even without magnetic fields. It came as a big surprise, however, that the polarized spectrum was as richly structured as the ordinary intensity spectrum but without resembling it, as if a new spectral face of the Sun had been unveiled, and we had to start over again to identify the various spectral structures and their physical origins. It was therefore natural to call this new and unfamiliar spectrum the ``Second Solar Spectrum'' (\cite{stenflo-ivanov91}; \cite{stenflo-sk97}). A spectral atlas has been produced, which in three volumes covers the Second Solar Spectrum from 3160 to 6995\,\AA\ (\cite{stenflo-gandorf00,stenflo-gandorf02,stenflo-gandorf05}). 

The Second Solar Spectrum exists as a fundamentally non-magnetic phenomenon, but it is modified by magnetic fields, it is the playground for the Hanle effect. Because of the rich structuring of the Second Solar Spectrum and the diverse behavior of the different spectral lines, it contains a variety of novel opportunities to diagnose solar magnetism in ways not possible with the Zeeman effect. Here we will only illustrate a few examples of this. Further details can be found in the proceedings of the series of Solar Polarization Workshops (\cite{stenflo-spw1}; \cite{stenflo-spw2}; \cite{stenflo-spw3}; \cite{stenflo-spw4}). 

\begin{figure}
 \includegraphics[angle=-90.,width=1.\textwidth]{\figspath/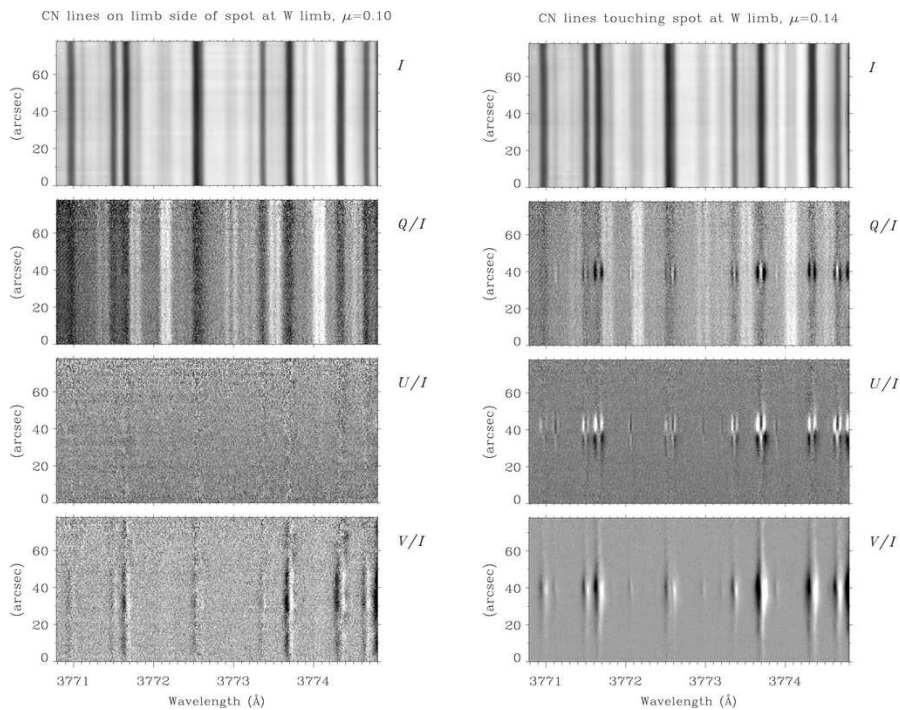}
\centering
  \caption[]{\label{stenflo-fig:CN}Molecular CN lines in the Second Solar Spectrum (the bright bands in Stokes $Q/I$). Note the absence of scattering polarization in $U/I$ and significant variation of $Q/I$ along the slit, in contrast to the surrounding atomic lines, which show the familiar signatures of the transverse and longitudinal Zeeman effects. The recording was made with ZIMPOL at Kitt Peak at $\mu=0.1$ inside the west solar limb (\cite{stenflo-s07}).}
\end{figure}

The structuring in the Second Solar Spectrum is governed by previously unfamiliar physical processes, like quantum interference between atomic levels, hyperfine structure and isotope effects, optical pumping, molecular scattering, and enigmatic, as yet unexplained phenomena that appear to defy quantum mechanics as we know it (cf.\ \cite{stenflo-s04}). The identification and interpretation of the various polarized structures have presented us with fascinating theoretical challenges, and we have now reached a good qualitative understanding of the underlying physics in most but not all of the cases. Here we will limit ourselves to illustrate the case of molecular scattering. 

The spectral Stokes vector images (intensity $I$, linear polarizations $Q/I$ and $U/I$, circular polarization $V/I$) in Fig.~\ref{stenflo-fig:CN} illustrate the behaviour of scattering polarization in the CN molecular lines in the wavelength range 3771--3775\,\AA, in solar regions of different degrees of magnetic activity. The CN lines have the appearance of emission lines in $Q/I$ with little if any spatial variations along the spectrograph slit, in contrast to the surrounding atomic lines, which exhibit the characteristic signatures of the transverse Zeeman effect. This would seem to imply that the molecular lines are not affected by magnetic fields, since we see no spatial structuring due to the Hanle effect, in contrast to the chromospheric Ca\,{\sc i} 4227\,\AA\ line in Fig.~\ref{stenflo-fig:hanle2}, where we see dramatic $Q/I$ and $U/I$ variations along the slit due to the Hanle effect. A careful analysis of the observed $Q/I$ amplitudes in the molecular lines reveal however, that they are indeed affected (depolarized) by the Hanle effect, but by a magnetic field that is tangled and structured on subresolution scales, and therefore does not show resolved variations along the slit or any $U/I$ signatures (Hanle rotation). 

The model dependence in the translation of polarization amplitudes to field strengths can be suppressed by using combinations of spectral lines that behave similarly in all respects except for their sensitivity to the Hanle effect. This {\it differential Hanle effect} (\cite{stenflo-setal98}) is similar to the line-ratio technique for the Zeeman effect that we discussed in Sect.~\ref{stenflo-sec:fluxtube}. Its effectiveness depends on our ability to find optimum line combinations that allow us to isolate the Hanle effect from all the other effects. It turns out to be much easier to find optimum line pairs among the molecular lines than among the atomic lines. This technique has been successfully used by \citet{stenflo-bf04} with a pair of C$_2$ molecular lines to determine the strength (15\,G) of the tangled or turbulent field. The molecular lines are found to give systematically lower field strengths than the atomic lines, which can be explained in terms of spatial structuring of the turbulent field on the granulation scale (\cite{stenflo-trujetal04}). 3-D radiative transfer modelling shows that the molecular abundance is highest inside the granules, which implies that the turbulent field is preferentially located in the intergranular lanes while containing structuring that continues far below the granulation scales. In the next section we will consider how far down this structuring is expected to continue. 


\section{Scale spectrum of the magnetic structures}\label{sec:scales}
Magnetic fields permeate the Sun with its convection zone. The turbulent convection, which penetrates into the photosphere, tangles the frozen-in magnetic field lines and thereby structures the field on a vast range of scales. The structuring continues to ever smaller scales, until we reach the scales where the frozen-in condition ceases to be valid and the field decouples from the turbulent plasma. This happens when the time scale of magnetic diffusion becomes shorter than the time scale of convective transport. The ratio between these two time scales is represented by the magnetic Reynolds number 
\begin{equation}
R_m =\mu_0\,\sigma\,\ell_c\,\,v_c
\end{equation}
in SI units. $\sigma$ is the electrical conductivity, $\ell_c$ the characteristic length scale, $v_c$ the characteristic velocity. $\mu_0=4\pi\,\,10^{-7}$. For large scales, when $R_m\gg 1$, the field lines are effectively frozen in and carried around by the convective motions. For sufficiently small scales $R_m\ll 1$, the field decouples and diffuses through the plasma. The end of the scale spectrum is where the decoupling occurs, namely where $R_m\approx 1$. 

To calculate the decoupling scale we need to know how the characteristic turbulent velocity $v_c$ scales with $\ell_c$. Such a scaling law is given in the Kolmogorov theory of isotropic turbulence. In the for us relevant inertial range it is 
\begin{equation}
v_c=k\,\,\ell_c^{1/3}\,,
\end{equation}
where $k$ is a constant. An estimate of $k\approx 25$ can be obtained from the observed properties of solar granulation (\AA ke Nordlund, private communication). 

Note that this type of scaling should apply to the photosphere in spite of its stratification, since the inertial range that we are considering occurs at scales much smaller than the photospheric scale height. This small-scale turbulence does not ``feel'' the stratification and is therefore nearly isotropic, in contrast to the larger scales. 

For $R_m=1$, these two equations give us the diffusion scale 
\begin{equation}
\ell_{\rm diff}=1/(\mu_0\,\sigma\,k)^{3/4}\,.
\end{equation}
Inserting the Spitzer conductivity in SI units,   
\begin{equation}
\sigma=10^{-3}\,\,T^{3/2}\,,
\end{equation}
we obtain 
\begin{equation}
\ell_{\rm diff}=5\times 10^5\,/\,T^{9/8}\,.
\end{equation}
For $T=10^4\,$K (a rounded value that is representative of the lowest part of the photosphere or upper boundary of the convection zone), $\ell_{\rm diff}\approx15\,$m. 

Note that the ordinary, non-magnetic Reynolds number is still very high at these 10\,m scales. Thus the turbulent spectrum continues to much smaller scales down to the viscous diffusion limit, but without contributing to magnetic structuring at these scales. 

The present-day spatial resolution limit in solar observations lies around 100\,km. This is 4 orders of magnitude larger than the smallest magnetic structures that we can expect. Therefore, in spite of conspicuous advances in high-resolution imaging, much of the structuring will remain unresolved in any foreseeable future.


\section{Beyond the standard model: scaling laws and PDFs for a fractal-like field}\label{stenflo-sec:fractal}
Time has come to replace the previous dualistic magnetic-field
paradigm or two-component ``standard model'' with a scenario
characterized by probability density functions (PDFs). While the
strong-field tail of such a distribution corresponds to the ``flux
tubes'' of the standard model, the bulk of the PDF corresponds to the
``turbulent field'' component. Instead of using two different
interpretational models for the Zeeman and Hanle effects when
diagnosing the spatially unresolved domain, it is more logical to
apply a single, unified interpretational model based on PDFs for both
these effects. The diagnostic tools for this unified and much more
realistic approach are currently being developed (cf.\
\cite{stenflo-sampoorna_pdf08}, \cite{stenflo-sampoorna2009}).

\begin{figure}
 \includegraphics[angle=-90.,width=1.\textwidth]{\figspath/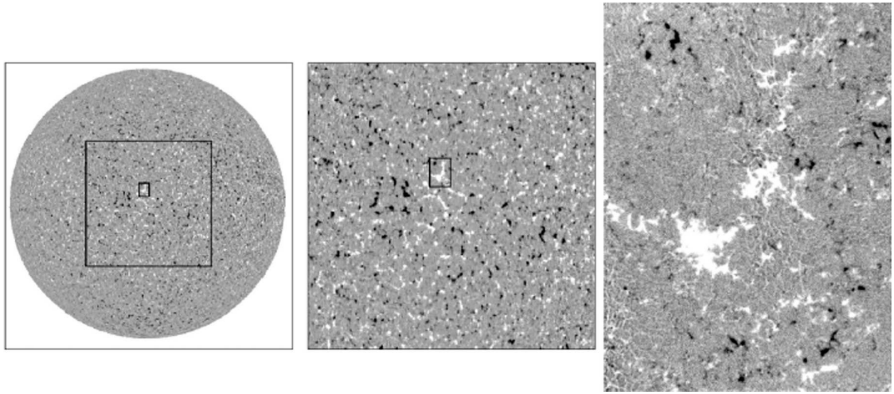}
  \caption[]{\label{stenflo-fig:fractal}The fractal-like nature of quiet-sun magnetic fields becomes apparent as we zoom in on ever smaller scales. The two left maps are from a Kitt Peak full-disk magnetogram of 9 February 1996, while the right, high-resolution magnetogram was recorded on the same day at the Swedish La Palma Observatory (courtesy G\"oran Scharmer). The La Palma magnetogram covers an area that is only 0.35\,\%\ of the map next to it.}
\end{figure}

This task is complicated because there are PDFs for both field strength and field orientation, and they appear to vary spatially on the granulation scale, as suggested by the different Hanle behavior of atomic and molecular lines. To clarify this we need to resolve the solar granulation in Hanle effect observations. Furthermore we know much less about the PDF for the angular field distribution than we know about the PDF for the vertical field strengths. For theoretical reasons we expect the angular and strength distributions to be coupled to each other. Strong fields are more affected by buoyancy forces, which make the angular distribution more peaked around the vertical direction. Small-scale, weak fields on the other hand are passively tangled by the turbulent motions and are therefore expected to have a more isotropic distribution. The issue is confused by the recent Hinode finding that there appears to be substantially more horizontal than vertical magnetic flux on the quiet Sun (\cite{stenflo-litesetal08}), which finds support in some numerical simulations (\cite{stenflo-sv08}). The implications of these findings for the angular PDFs have not yet been clarified. 

Another fundamental issue is the dependence of these various PDFs on scale size. To wisely select the interpretational models to be used to diagnose the unresolved domain we need to understand the relevant scaling laws. Explorations of the magnetic-field pattern in magnetograms (the spatially resolved domain) and in numerical simulations indicate a high degree of self-similarity and fractal-like behavior. This would justify the use of PDF shapes that are found from the resolved domain to be applied to diagnostics of the unresolved domain. On the other hand there are reasons to expect possible deviations from such scale invariance. We have already seen indications for a difference between the PDFs in granules and in intergranular lanes. The current spatial resolution limit (about 100\,km) also marks the boundary between optically thick and thin elements, as well as between elements governed by the atmospheric stratification effects (scale height) and elements that are too small to ``feel'' this stratification. The 100\,km scale is therefore expected to be of physical significance and may influence the behavior of the scaling laws. 

The fractal nature of the field is illustrated in Fig.~\ref{stenflo-fig:fractal} as we zoom in on the quiet-sun magnetic pattern at the center of the solar disk. There is a coexistence of weak and strong fields over a wide dynamic range. The PDF for the vertical field-strength component is nearly scale invariant and can be well represented by a Voigt function with a narrow Gaussian core and ``damping wings'' extending to kG values (\cite{stenflo-sh02,stenflo-sh03}). A fractal dimension of 1.4 has been found from both observations and numerical simulations (\cite{stenflo-jetal03}). The simulations indicate that this fractal behavior extends well into the spatially unresolved domain. 

The Second Solar Spectrum opens a new window to explorations of previously inaccessible aspects of solar magnetism. With the vast amounts of hidden magnetic energy in the spatially unresolved magnetoconvective spectrum, the determination of the properties of the hidden field is a central task for contemporary solar physics. The quality of the determination depends on the interpretational models that we choose. For an optimum choice we need to understand the scaling laws of the fractal-like magnetic field and the role of various physical scales that may cause deviations from scale invariance of the pattern. Insight into the scaling behavior can be advanced with improved spatial resolution of the observations, combined with guidance from numerical simulations.

\begin{small}



\end{small}


\end{document}